\documentclass[a4paper,10pt]{article}
\pdfoutput=1

\usepackage{hyperref}
\usepackage{gensymb}  
\usepackage{
graphicx,
hyperref,
amsmath,
amssymb,
charter,
xcolor,
ifluatex,
multirow}
        
\definecolor{Gray}{gray}{0.95}
\definecolor{RGray}{gray}{0.85}
\definecolor{CGray}{gray}{0.92}

\definecolor{tit}{rgb}{0.1,0.2,0.4}
\definecolor{blus}{cmyk}{1,1,0,0.6}
\definecolor{verde}{cmyk}{0.92,0,0.59,0.25}

\usepackage{tipa}
\usepackage{amsmath,amssymb,amsfonts, bm}
\usepackage{epsfig}
\usepackage{graphicx}
\usepackage{slashed}
\usepackage{color}

\usepackage{caption}
\usepackage{subcaption}
\usepackage{slashed} 
\usepackage{float}

\newcommand{\D}{{\cal D}}
\newcommand{\U}{{\cal U}}

\newcommand{\M}{{\cal M}}

\usepackage{calc}

\newcommand{\be}{\begin{equation}}
\newcommand{\ee}{\end{equation}}

\newcommand{\bea}{\begin{eqnarray}}
\newcommand{\eea}{\end{eqnarray}}

\newcommand{\bfig}{\begin{figure}}
\newcommand{\efig}{\end{figure}}

\makeatletter
\newcommand*{\rom}[1]{\expandafter\@slowromancap\romannumeral #1@}
\makeatother

\begin{document}
\allowdisplaybreaks
\vspace*{-2.5cm}
\begin{flushright}
{\small
IIT-BHU
}
\end{flushright}

\vspace{2cm}

\begin{center}
{\LARGE \bf \color{tit} Standard HVM }\\[1cm]

{\large\bf Gauhar Abbas\footnote{email: gauhar.phy@iitbhu.ac.in} \\  [7mm]
{\it  } {\em Department of Physics, \\ 
\vspace{0.5cm}
Indian Institute of Technology (BHU),\\ 
\vspace{0.5cm}
Varanasi 221005, India}}\\[3mm]

\vspace{1cm}
{\large\bf\color{blus} Abstract}
\begin{quote}
We discuss a standard hierarchical VEVs model which predicts the leptonic mixing angles in terms  of the Cabibbo angle, and masses of strange and charm quarks as $\sin \theta_{12}^\ell   \geq 1 - 2 \sin \theta_{12}$,  $\sin \theta_{23}^\ell   \geq 1 -  \sin \theta_{12}$,  and $\sin \theta_{13}^\ell  \geq \sin \theta_{12} - \frac{m_s}{m_c}$ for the normal mass ordering of neutrinos.  This results in  very precise predictions of the leptonic mixing angles given by  $\sin \theta_{12}^\ell =  0.55 \pm 0.00134 $,  $\sin \theta_{23}^\ell =  0.775 \pm 0.00067 $,  and   $\sin \theta_{13}^\ell = 0.1413 - 0.1509 $.   Furthermore, we predict neutrinos to be the Dirac kind disfavouring the inverted mass hierarchy.  The standard hierarchical VEVs model predicts a possible new class of the dark matter candidate, named as neutrinic dark matter.
\end{quote}

\thispagestyle{empty}
\end{center}

\begin{quote}
{\large\noindent\color{blus} 
}

\end{quote}

\newpage
\setcounter{footnote}{0}
\section{Introduction}
The origin of the flavour of the standard model (SM) is a fascinating problem that may lead to  physics beyond the SM. This problem demands an explanation   for  the fermionic mass spectrum and mixing of the SM.  There exist  different ways to solve this problem.  For instance,  a technicolour framework  can solve this problem through the   multi-fermion chiral condensates of a dark-technicolor (DTC) symmetry \cite{Abbas:2017vws,Abbas:2020frs}.     An  Abelian flavour symmetry can be used as well~\cite{Froggatt:1978nt,flavor_symm1,Chun:1996xv,flavor_symm2,flavor_symm3,Davidson:1983fy,Davidson:1987tr,Berezhiani:1990wn,Berezhiani:1990jj,Berezhiani:1989fp,Sakharov:1994pr}.  We can also have a solution based on   loop-suppressed couplings to the Higgs~\cite{higgs_coup}, wave-function localization scenario~\cite{wf_local}, the  compositeness~\cite{partial_comp}, extra-dimensional framework \cite{Fuentes-Martin:2022xnb},  and    discrete symmetries\cite{Abbas:2018lga,Hinata:2020cdt,Abbas:2022zfb,Higaki:2019ojq,Abbas:2024dfh}, and flavour deconstruction \cite{Bordone:2017bld}-\cite{Greljo:2024ovt}.

Hierarchical VEVs model (HVM) redefines the flavour problem of the SM  in terms of the VEVs hierarchy,  and provide an explanation to the quark mixing angles in terms of the ratios of the hierarchical VEVs \cite{Abbas:2017vws}.   In the ultra-violet completion (UV) of  HVM,  the hierarchical VEVs are  chiral multi-fermion condensates of a DTC dynamics depending only on a single parameter $\Lambda_{\rm DTC}$, i.e.,  the scale  of the underlying non-Ableian DTC gauge theory\cite{Abbas:2020frs}.  For a recent review of the HVM,  see reference \cite{Abbas:2023ivi}.

In the simplest version of  the HVM,  the SM is  subjected to  the imposition of the three Abelian discrete symmetries  $\mathcal{Z}_2$, $\mathcal{Z}_2^\prime$ an $\mathcal{Z}_2^{\prime \prime} $,  and charged fermionic mass spectrum is explained in terms of the VEVs of the six gauge singlet scalar fields $\chi _i $, assumed to be the bound states of a new strong dynamics,  through the dimension-5 operators  \cite{Abbas:2017vws}.  In the HVM, the neutrino masses are recovered by adding an additional gauge singlet scalar field  $\chi _7 $ through the type-\rom{1} seesaw mechanism\cite{seesaw}.  However,  no predictions for the neutrino mixing angles were made.

Predicting neutrino mixing angles is one of the most difficult challenges beyond the SM.   For a review of neutrino models,   see reference \cite{Valle}.  The reason is that the observed pattern of the neutrino mixing angles is entirely different  from that of the quark mixing angles.  For instance,  a global fit to neutrino oscillation data provides the following values of the neutrino oscillation parameters for the normal mass ordering \cite{deSalas:2020pgw},
\bea
\Delta m_{21}^2 &=& (7.50^{+0.64}_{-0.56}) \times 10^{-5} {\rm eV}^2, ~|\Delta m_{31}^2| = (2.55\pm 0.08) \times 10^{-3} \rm{eV}^2,  \\ \nonumber
 \sin \theta_{12}^\ell    &=&  (0.564_{-0.043}^{+0.044}),~
\sin  \theta_{23}^\ell  =   (0.758_{-0.099}^{+0.023}),~   \sin \theta_{13}^\ell  =  (0.1483_{-0.0069}^{+0.0067}),
\eea
where the errors are in 3 $\sigma$ range.

We observe that only $\sin \theta_{13}^\ell $ is closer to the Cabibbo angle $\sin \theta_{12}^q$, and $\sin \theta_{12}^q-\sin \theta_{13}^\ell \approx 0.06923 - 0.08427$.  Furthermore, although the mixing angle  $\sin \theta_{23}^\ell$ is of the order one,  it is not exactly one.  Moreover, the $\sin \theta_{12}^\ell $ is large, however, it cannot be considered  order one.

On the other side, the quark mixing angles are \cite{pdg22},   
\bea
\label{Qangle}
\sin \theta_{12}    &=&  (0.225 \pm 0.00067),~
\sin \theta_{23}  =  (0.04182^{+0.00085}_{-0.00074}),~  \sin \theta_{13}  =  (0.00369 \pm 0.00011).
\eea
In the case of quark-mixing, $\sin \theta_{23} \approx \sin \theta_{12}^2 $ and remarkable observation is $\sin \theta_{13} < \sin \theta_{12}^3 $.

The really important and challenging question is, if these observations can be predicted from a fundamental theory. In this work,   we shall discuss what we refer to as the standard HVM (SHVM), and show that the above  observations can be explained in the SHVM.

We shall show that the SHVM leads to a  multiple axion-like particles (ALPs) scenario. ALPs can play the role of cold dark matter \cite{Preskill:1982cy,Abbott:1982af,Dine:1982ah,Marsh:2015xka,Chadha-Day:2021szb}, and can address the strong CP problem \cite{Peccei:1977hh,Wilczek:1977pj}.  Due to the multiple ALPs scenario, the SHVM may be a consequence of a string type dynamics.  This is because of the existence of an axiverse, that is, emergence of  a large number of ALPs in string theory \cite{Svrcek:2006yi,Cicoli:2012sz,Broeckel:2021dpz,Demirtas:2021gsq,Gendler:2023kjt}.

A field theory  axiverse ($\pi$ axiverse ) is an alternative to string theory  axiverse, which also produces a large number of ALPs \cite{Alexander:2024nvi}. The field theory  axiverse  is based on a confining gauge theory referred as dark QCD \cite{Bai:2013xga,Tsai:2020vpi,Alexander:2020wpm}. There are  $N_f^2-1$ axionlike states  in dark QCD \cite{Alexander:2024nvi}. 

The UV completion of the SHVM can come from   three strong QCD-like gauge dynamics, namely, $SU(N)_{TC}$, $SU(N)_{DTC}$, and $SU(N)_{F}$, where, TC stands for technicolor, DTC is for dark-technicolor, and F represents the strong dynamics of vector-like fermions of a dark-QCD symmetry \cite{Abbas:2020frs}.  Therefore, emergence of ALPs is natural in the SHVM from its UV dynamics.  Moreover, the  dark-technicolor paradigm is a good example of the field theory  axiverse.

The most important fact in support of the SHVM and its UV completion, the  dark-technicolor paradigm, comes from a recent development where it is shown that a conformal strong dynamics  can provide the eletroweak symmetry breaking  \cite{Chatterjee:2024dgw}.   This dynamics in the case of the dark-techncicolour paradigm is the $SU(N)_{TC}$ symmetry, which generates the SM Higgs field.  Hence, the SHVM with  dark-technicolor paradigm as a UV completion, is an interesting scenario to address the flavour origin of the SM.

\section{The standard HVM}
\label{sec2}
The standard HVM is achieved by enlarging the discrete symmetries used in HVM. We  impose a generic   $\mathcal{Z}_{\rm N} \times \mathcal{Z}_{\rm M} \times \mathcal{Z}_{\rm P}$ flavour symmetry on the SM.  This symmetry naturally emerges through the breaking of three axial $U(1)$ groups in the UV completion of HVM  \cite{Abbas:2020frs,Abbas:2023ivi}. For the sake of completeness, we present an explicit UV construction of the SHVM in appendix where the  gauge singlet scalar fields $\chi _i $  and the $\mathcal{Z}_{\rm N} \times \mathcal{Z}_{\rm M} \times \mathcal{Z}_{\rm P}$ flavour symmetry naturally arise from a DTC dynamics.

The transformations of the composite scalar fields  $\chi _i $  under the $\mathcal{G}_{\rm SM} \equiv SU(3)_c \times SU(2)_L \times U(1)_Y$ symmetry of the SM are given as,
\begin{eqnarray}
\chi _i :(1,1,0),
 \end{eqnarray} 
where $i=1-6$.

The masses of the charged fermions are  produced by the  dimension-5 operators  written as,
\bea
\label{mass2}
{\mathcal{L}} &=& \dfrac{1}{\Lambda }\Bigl[  y_{ij}^u  \bar{\psi}_{L_i}^{q}  \tilde{\varphi} \psi_{R_i}^{u}   \chi _i +     
   y_{ij}^d  \bar{\psi}_{L_i}^{q}   \varphi \psi_{R_i}^{d}  \chi _{i}   +   y_{ij}^\ell  \bar{\psi}_{L_i}^{\ell}   \varphi \psi_{R_i}^{\ell}  \chi _{i} \Bigr]  
+  {\rm H.c.},
\eea
$i$ and $j$   are generation indices,  $ \psi_{L}^q,  \psi_{L}^\ell  $ are  the  quark and leptonic doublets,  $ \psi_{R}^u,  \psi_{R}^d, \psi_{R}^\ell$ denote the right-handed up,  down-type  quarks and  leptons,  $\varphi$ and $ \tilde{\varphi}= -i \sigma_2 \varphi^* $  represent the SM Higgs field, and its conjugate and $\sigma_2$ is  the second Pauli matrix.

We produce the required pattern of the charged fermions masses for the SHVM by assigning the following generic charges  under the   $\mathcal{Z}_{\rm N} \times \mathcal{Z}_{\rm M} \times \mathcal{Z}_{\rm P}$ flavour symmetry.
\begin{align}
\psi_{L_1}^{q} &: (+, 1, \omega^{\rm P-1}_{14}),~ \psi_{L_2}^{q}: (+, 1, \omega_{14}),~  \psi_{L_3}^{q}: (+, 1, \omega^{ 2}_{14}), \\ \nonumber
u_{R} &: (-, 1, \omega^{\rm P-3}_{14}),~ c_{R}: (+, 1, \omega^6_{14}),~  t_{R}: (+, 1, \omega^{ 4}_{14}), \\ \nonumber
\psi_{L_1}^{\ell} &: (+, \omega^{3}_{4}, \omega^{12}_{14}),~ \psi_{L_2}^{\ell}: (+, \omega^{3}_4, \omega^{10}_{14}),~  \psi_{L_3}^{\ell}: (+, \omega^{3}_4, \omega^{6}_{14}), \\ \nonumber
e_{R} &: (-, \omega^{3}_4, \omega^{10}_{14}),~ \mu_{R}: (+, \omega^{3}_4, \omega^{P-1}_{14}),~  \tau_{R}: (+, \omega^{3}_4, \omega_{14}), \\ \nonumber
\chi_1 &: (-, 1, \omega^{2}_{14}),~ \chi_2: (+, 1, \omega^{5}_{14}),~  \chi_3: (+, 1, \omega^2_{14}), \\ \nonumber
\chi_4 &: (+, 1, \omega^{13}_{14}),~ \chi_5: (+, 1, \omega^{11}_{14}),~  \chi_6: (+, 1, \omega^6_{14}), 
\end{align}
where $\omega_4$ is the fourth  and  $\omega_{14}$ is the fourteenth root of unity corresponding to the symmetries $\mathcal{Z}_4$  and $\mathcal{Z}_{14}$,  respectively.  Moreover, $\rm N= 2$, $\rm M\geq 4$,  and $\rm P \geq 14$ are needed.  For recovering the desired flavour structure of the charged fermions in the standard HVM,  we must have  $\rm N= 2$.

It turns out that  for the flavour structure of the standard HVM assuming normal neutrino mass ordering,   we must have $\rm N= 2$, $\rm M\geq 4$,  and $\rm P \geq 14$.  For instance, we use the $\mathcal{Z}_2 \times \mathcal{Z}_4 \times \mathcal{Z}_{14} $ symmetry, and assign charges to different fermionic and scalar fields under this symmetry,  as shown in table   \ref{tab1}. 

 \begin{table}[ht]
\begin{center}
\begin{tabular}{|c|c|c|c||c|c|c|c||c|c|c|c||c|c|c|c|c|}
  \hline
  Fields                               &   $\mathcal{Z}_2$  &  $\mathcal{Z}_4$   &  $\mathcal{Z}_{14}$   & Fields   &  $\mathcal{Z}_2$   &  $\mathcal{Z}_4$ &  $\mathcal{Z}_{14}$ & Fields   & $\mathcal{Z}_2$  & $\mathcal{Z}_4$  &  $\mathcal{Z}_{14}$  & Fields  &  $\mathcal{Z}_2$   &  $\mathcal{Z}_4$  &  $\mathcal{Z}_{14}$ \\
  \hline
 $u_{R}$                        &     -   &     $ 1$          &    $\omega^{11}_{14}$ & $d_{R} $, $ s_{R}$, $b_{R}$    &     +    & $ 1$        &    $\omega^{12}_{14}$  & $ \psi_{L_3}^{q} $       &    +     &  $ 1$    &   $\omega^2_{14}$   &  $\tau_R$      &   +  &  $\omega^3_{4}$      &     $\omega_{14}$             \\
  $c_{R}$                       &     +   &    $ 1$          &    $\omega^6_{14}$   &  $\chi _4$                         &      +  &  $ 1$      &  $\omega^{13}_{14}$  &  $ \psi_{L_1}^\ell $                          &     +   & $\omega^3_4$     &  $\omega^{12}_{14}$                          &   $\nu_{e_R}$   &    +   &   $\omega_4$     &      $\omega^{8}_{14}$        \\
   $t_{R}$                        &     +   &    $1$         &    $\omega^{4}_{14}$   & $\chi _5$                         &      +  &  $1$    &  $\omega^{11}_{14}$  &  $ \psi_{L_2}^{\ell} $     &      +  & $\omega^3_4$      &  $\omega^{10}_{14}$     & $\nu_{\mu_R}$                   &     -  &   $\omega_4$    &   $\omega^3_{14}$                    \\
  $\chi _1$                        &      -  &   $ 1 $      &    $\omega^2_{14}$    &   $\chi _6$                          &      +  &  $ 1$      &   $ \omega^{10}_{14}$    &   $ \psi_{L_3}^{ \ell} $       &    +     &  $\omega^3_4$    &   $\omega^6_{14}$                                 & $\nu_{\tau_R}$                    &     -   &  $\omega_4$     &   $\omega^3_{14}$          \\
  $\chi _2$                   & +     &       $ 1$      &  $\omega^5_{14}$   & $ \psi_{L_1}^q $                          &      +  &  $1$      &  $\omega^{13}_{14}$  & $e_R$    &      -   &   $\omega^3_4$       &    $\omega^{10}_{14}$      &  $\chi_7 $                          &      -   &  $\omega^2_4$   &     $\omega^8_{14}$                                               \\
   $\chi _3$                  &    +   &       $ 1$    & $ \omega^2_{14}$        & $ \psi_{L_2}^{q} $     &      +  & $1 $      &  $\omega_{14}$  &   $ \mu_R$     &   +  & $\omega^3_4$       &     $\omega^{13}_{14}$      &  $ \varphi $                           &      +  &1     &   1                                            \\
  \hline
     \end{tabular}
\end{center}
\caption{The charges of left- and right-handed fermions  and  scalar fields under $\mathcal{Z}_2$, $\mathcal{Z}_4$  and $\mathcal{Z}_{14}$ symmetries for the normal mass ordering. The $\omega_4$ is the fourth  and  $\omega_{14}$ is the fourteenth root of unity corresponding to the symmetries $\mathcal{Z}_4$  and $\mathcal{Z}_{14}$,  respectively.}
 \label{tab1}
\end{table} 

The Lagrangian providing masses to charged fermions is now given as,
\bea
\label{mass2}
{\mathcal{L}_{f}} &=& \dfrac{1}{\Lambda }\Bigl[  y_{11}^u  \bar{\psi}_{L_1}^{q}  \tilde{\varphi} \psi_{R_1}^{u}   \chi _1 +  y_{13}^u  \bar{\psi}_{L_1}^{q}  \tilde{\varphi} \psi_{R_3}^{u}   \chi _2^\dagger  +  y_{22}^u  \bar{\psi}_{L_2}^{q}  \tilde{\varphi} \psi_{R_2}^{u}   \chi _2^\dagger  +  y_{23}^u  \bar{\psi}_{L_2}^{q}  \tilde{\varphi} \psi_{R_3}^{u}   \chi _5  +  y_{32}^u  \bar{\psi}_{L_3}^{q}  \tilde{\varphi} \psi_{R_2}^{u}   \chi _6 \\ \nonumber
&&  +  y_{33}^u  \bar{\psi}_{L_3}^{q}  \tilde{\varphi} \psi_{R_3}^{u}   \chi _3^\dagger +     
   y_{11}^d  \bar{\psi}_{L_1}^{q}   \varphi \psi_{R_1}^{d}  \chi_{4}^\dagger +     
   y_{12}^d  \bar{\psi}_{L_1}^{q}   \varphi \psi_{R_2}^{d}  \chi_{4}^\dagger +     
   y_{13}^d  \bar{\psi}_{L_1}^{q}   \varphi \psi_{R_3}^{d}  \chi_{4}^\dagger +     
   y_{21}^d  \bar{\psi}_{L_2}^{q}   \varphi \psi_{R_1}^{d}  \chi_{5}^\dagger   \\ \nonumber
   && + y_{22}^d  \bar{\psi}_{L_2}^{q}   \varphi \psi_{R_2}^{d}  \chi_{5}^\dagger + y_{23}^d  \bar{\psi}_{L_2}^{q}   \varphi \psi_{R_3}^{d}  \chi_{5}^\dagger + y_{31}^d  \bar{\psi}_{L_3}^{q}   \varphi \psi_{R_1}^{d}  \chi_{6}^\dagger  + y_{32}^d  \bar{\psi}_{L_3}^{q}   \varphi \psi_{R_2}^{d}  \chi_{6}^\dagger + y_{33}^d  \bar{\psi}_{L_3}^{q}   \varphi \psi_{R_3}^{d}  \chi_{6}^\dagger  \\ \nonumber
   && +   y_{11}^\ell  \bar{\psi}_{L_1}^{\ell}   \varphi \psi_{R_1}^{\ell}  \chi _{1}  +   y_{12}^\ell  \bar{\psi}_{L_1}^{\ell}   \varphi \psi_{R_2}^{\ell}  \chi _{4}  +   y_{13}^\ell  \bar{\psi}_{L_1}^{\ell}   \varphi \psi_{R_3}^{\ell}  \chi _{5}  +   y_{22}^\ell  \bar{\psi}_{L_2}^{\ell}   \varphi \psi_{R_2}^{\ell}  \chi _{5}  +   y_{23}^\ell  \bar{\psi}_{L_2}^{\ell}   \varphi \psi_{R_3}^{\ell}  \chi _{2}^\dagger  \\ \nonumber
&& +   y_{33}^\ell  \bar{\psi}_{L_3}^{\ell}   \varphi \psi_{R_3}^{\ell}  \chi _{2} +  {\rm H.c.} \Bigr].
\eea

The fermionic mass spectrum can be  explained in terms of  the   VEVs pattern    $ \langle \chi _4 \rangle > \langle \chi _1 \rangle $, $ \langle \chi _2 \rangle >> \langle \chi _5 \rangle $, $ \langle \chi _3 \rangle >> \langle \chi _6 \rangle $, $ \langle \chi _{3} \rangle >> \langle \chi _{2} \rangle >> \langle \chi _{1} \rangle $, and  $ \langle \chi _6 \rangle >> \langle \chi _5 \rangle >> \langle \chi _4 \rangle $.  The mass matrices of up,  down quarks and leptons turn out to be,
\begin{align}
\label{mUD}
\M_\U & =   \dfrac{ v }{\sqrt{2}} 
\begin{pmatrix}
y_{11}^u  \epsilon_1 &  0  & y_{13}^u  \epsilon_2    \\
0    & y_{22}^u \epsilon_2  & y_{23}^u  \epsilon_5   \\
0   &  y_{32}^u  \epsilon_6    &  y_{33}^u  \epsilon_3 
\end{pmatrix},  
\M_\D = \dfrac{ v }{\sqrt{2}} 
 \begin{pmatrix}
  y_{11}^d \epsilon_4 &    y_{12}^d \epsilon_4 &  y_{13}^d \epsilon_4 \\
  y_{21}^d \epsilon_5 &     y_{22}^d \epsilon_5 &   y_{23}^d \epsilon_5\\
    y_{31}^d \epsilon_6 &     y_{32}^d \epsilon_6  &   y_{33}^d \epsilon_6\\
\end{pmatrix},
\M_\ell =\dfrac{ v }{\sqrt{2}} 
  \begin{pmatrix}
  y_{11}^\ell \epsilon_1 &    y_{12}^\ell \epsilon_4  &   y_{13}^\ell \epsilon_5 \\
 0 &    y_{22}^\ell \epsilon_5 &   y_{23}^\ell \epsilon_2\\
   0  &    0  &   y_{33}^\ell \epsilon_2 \\
\end{pmatrix},
\end{align} 
where $\epsilon_i = \dfrac{\langle \chi _{i} \rangle }{\Lambda}$ and  $\epsilon_i<1$.  

The masses of charged fermions  are,
\begin{eqnarray}
\label{mass1c}
m_t  &=& \ \left|y^u_{33} \right| \epsilon_3 v/\sqrt{2}, ~
m_c  = \   \left|y^u_{22} \epsilon_2  - \dfrac{y_{23}^u  y_{32}^u \epsilon_5 \epsilon_6 }{y_{33}^u \epsilon_3} \right|  v /\sqrt{2} ,~
m_u  =  |y_{11}^u  |\,  \epsilon_1 v /\sqrt{2},\nonumber \\
m_b  &\approx& \ |y^d_{33}| \epsilon_6 v/\sqrt{2}, 
m_s  \approx \   \left|y^d_{22} - \dfrac{y_{23}^d  y_{32}^d}{y_{33}^d} \right| \epsilon_5 v /\sqrt{2} ,\nonumber \\
m_d  &\approx&  \left|y_{11}^d - {y_{12}^d y_{21}^d \over  y^d_{22} - \dfrac{y_{23}^d  y_{32}^d}{y_{33}^d} }   -
{{y_{13}^d (y_{31}^d y_{22}^d - y_{21}^d y_{32}^d )-y_{31}^d  y_{12}^d  y_{23}^d } \over 
{ (y^d_{22} - \dfrac{y_{23}^d  y_{32}^d}{y_{33}^d})  y^d_{33}}}   \right|\,  \epsilon_4 v /\sqrt{2},\nonumber \\
m_\tau  &\approx& \ |y^\ell_{33}| \epsilon_2 v/\sqrt{2}, ~
m_\mu  \approx \   |y^\ell_{22} | \epsilon_5 v /\sqrt{2} ,~
m_e  =  |y_{11}^\ell   |\,  \epsilon_1 v /\sqrt{2}.
\end {eqnarray}
The quark mixing angles are \cite{tasi2000},
\begin{eqnarray}
\sin \theta_{12}  & \simeq&  \left|{y_{12}^d \epsilon_4 \over y_{22}^d \epsilon_5}   \right|= { \epsilon_4 \over \epsilon_5}, ~
\sin \theta_{23}  \simeq  \left| {y_{23}^d \epsilon_5  \over y_{33}^d \epsilon_6 }  - {y_{23}^u \epsilon_5  \over y_{33}^u \epsilon_3 }   \right|,~
\sin \theta_{13}  \simeq   \left|{y_{13}^d  \epsilon_4  \over y_{33}^d  \epsilon_6 } -{y_{13}^u  \epsilon_2  \over y_{33}^u  \epsilon_3 } \right|.
\end{eqnarray}

We assume all $|y_{ij}^{d}| =1 $, and write the quark mixing angles as,
\begin{eqnarray}
\sin \theta_{12}  & \simeq&   { \epsilon_{4} \over \epsilon_{5}}, ~ \\ \nonumber 
\sin \theta_{23} & \simeq &  { \epsilon_{5} \over \epsilon_{6}}, \\ \nonumber 
\sin \theta_{13}  &\simeq &  { \epsilon_{4} \over \epsilon_{6}} - \left|\frac{ y_{13}^u}{ y_{33}^u} \right| { \epsilon_{2} \over \epsilon_{3}}.
\end{eqnarray} 

The above  result leads to  the prediction of the quark mixing angle $ \theta_{13} $ given by,
\begin{eqnarray}
\sin \theta_{13}  \simeq   \sin \theta_{12}  \sin \theta_{23}  -  { \epsilon_{2} \over \epsilon_{3}}.
\end{eqnarray} 
for $|y_{ij}^{u}| =1 $.  The standard HVM is capable of predicting the  quark mixing angle $\sin \theta_{13}$ for $\epsilon_3= 0.55$, which is an allowed value of this parameter.

In general, the values of the $\epsilon_i$ for reproducing masses and mixing parameters are,
\begin{equation}
\label{epsi}
\epsilon_1 = 3.16 \times 10^{-6},~ \epsilon_2 = 0.0031,~ \epsilon_3 = 0.87,~\epsilon_4 = 0.000061,~\epsilon_5 = 0.000270,~\epsilon_6 = 0.0054,~\epsilon_7 = 7.18 \times 10^{-10}.  
\end{equation}

\section{Neutrino masses and mixing}
\label{sec3}
Now we address the issue of neutrino masses and mixing in the SHVM.  The neutrino masses are recovered by introducing the  three right-handed  neutrinos $\nu_{eR}$, $\nu_{\mu R}$, $\nu_{\tau R}$  to the SM. 

\subsection{Normal mass ordering}
We write the following Lagrangian to produce the neutrino masses,
\begin{eqnarray}
\label{massN}
-{\mathcal{L}}_{\rm Yukawa}^{\nu} &=&      y_{ij}^\nu \bar{ \psi}_{L_i}^\ell   \tilde{\varphi}  \nu_{f_R} \left[  \dfrac{ \chi_i \chi_j }{\Lambda^2} \right] +  {\rm H.c.}. 
\end{eqnarray}

The requirement that  the neutrino masses originates from  equation \ref{massN}  constrains   the symmetry  $\mathcal{Z}_{\rm P}$ allowing only $\rm P = 14$.  Thus,  $\rm P = 14$ is a magic number whose origin may lie in the UV theory.  The Lagrangian providing masses to neutrinos reads,
\bea
\label{mass2}
{\mathcal{L}_{f}} &=& \dfrac{1}{\Lambda^2 }\Bigl[   y_{11}^\nu  \bar{\psi}_{L_1}^{\ell}  \tilde{\varphi} \psi_{R_1}^{\nu}   \chi _1^\dagger \chi_7^\dagger +  y_{12}^\nu  \bar{\psi}_{L_1}^{\ell}  \tilde{\varphi} \psi_{R_2}^{\nu}   \chi _4^\dagger \chi_7 +  y_{13}^\nu  \bar{\psi}_{L_1}^{\ell}  \tilde{\varphi} \psi_{R_3}^{\nu}   \chi _4^\dagger \chi_7 +  y_{22}^\nu  \bar{\psi}_{L_2}^{\ell}  \tilde{\varphi} \psi_{R_2}^{\nu}   \chi _4 \chi_7 \\  \nonumber
&& +  y_{23}^\nu  \bar{\psi}_{L_2}^{\ell}  \tilde{\varphi} \psi_{R_3}^{\nu}   \chi _4 \chi_7 +  y_{32}^\nu  \bar{\psi}_{L_3}^{\ell}  \tilde{\varphi} \psi_{R_2}^{\nu}   \chi_5 \chi_7 +  y_{33}^\nu  \bar{\psi}_{L_3}^{\ell}  \tilde{\varphi} \psi_{R_3}^{\nu}   \chi_5 \chi_7 +  {\rm H.c.} \Bigr].
\eea

The Dirac mass matrix for neutrinos reads,
\begin{equation}
\label{NM}
\M_{\D} = \dfrac{v}{\sqrt{2}}  
\begin{pmatrix}
y_{11}^\nu   \epsilon_1 \epsilon_7   &  y_{12}^\nu   \epsilon_4 \epsilon_7  & y_{13}^\nu  \epsilon_4  \epsilon_7 \\
0   & y_{22}^\nu  \epsilon_4  \epsilon_7 &  y_{23}^\nu  \epsilon_4  \epsilon_7 \\
0   &   y_{32}^\nu  \epsilon_5  \epsilon_7   &  y_{33}^\nu  \epsilon_5  \epsilon_7
\end{pmatrix},
\end{equation}
and neutrino  masses can be written as,
\begin{eqnarray}
\label{mass1a}
m_3  &\approx&  |y^\nu_{33}|  \epsilon_5 \epsilon_7 v/\sqrt{2}, 
m_2  \approx     \left|y^\nu_{22} - \dfrac{y_{23}^\nu  y_{32}^\nu}{y_{33}^\nu} \right|  \epsilon_4 \epsilon_7 v /\sqrt{2},
m_1  \approx  |y_{11}^\nu  |\,  \epsilon_1 \epsilon_7 v /\sqrt{2}.
\end {eqnarray}
The masses of neutrinos turn out to be $\{m_3,m_2,m_1\} = \{5.05 \times 10^{-2}, 8.67 \times 10^{-3}, 2.67 \times 10^{-4} \}\, \text{eV}$
for  the values of $y_{ij}^\nu$ given in the appendix.

The remarkable predictions are   the neutrino mixing angles \cite{tasi2000},
\begin{eqnarray}
\sin \theta_{12}^\ell  &\simeq& \left|{y_{12}^\ell \epsilon_4  \over y_{22}^\ell  \epsilon_5}-{y_{12}^\nu  \over y_{22}^\nu }   +  {y_{23}^{\ell *} y_{13}^\nu \epsilon_4  \over y_{33}^\ell y_{33}^\nu  \epsilon_5}   \right| , 
\sin \theta_{23}^\ell  \simeq  \left|{y_{23}^\ell   \over y_{33}^\ell  }    -{y_{23}^\nu \epsilon_4  \over y_{33}^\nu  \epsilon_5}   \right| ,~
\sin \theta_{13}^\ell    \simeq \left|  {y_{13}^\ell   \epsilon_5  \over y_{33}^\ell  \epsilon_2 }   -{y_{13}^\nu \epsilon_4  \over y_{33}^\nu  \epsilon_5}   \right| .
\end{eqnarray}  
Assuming all the couplings of the order one,   we further predict ,
\begin{eqnarray}
\sin \theta_{12}^\ell  &\simeq&  \left|-{y_{12}^\nu  \over y_{22}^\nu } +{y_{12}^\ell \epsilon_4  \over y_{22}^\ell  \epsilon_5}+  {y_{23}^{\ell *} y_{13}^\nu \epsilon_4  \over y_{33}^\ell y_{33}^\nu  \epsilon_5}   \right| \geq   \left|-{y_{12}^\nu  \over y_{22}^\nu }   \right| -   \left| {y_{12}^\ell   \over y_{22}^\ell  } +  {y_{23}^{\ell *} y_{13}^\nu   \over y_{33}^\ell y_{33}^\nu  }  \right|  {\epsilon_4  \over   \epsilon_5} \approx  1 - 2 \sin \theta_{12}, \\ \nonumber
\sin \theta_{23}^\ell  &\simeq&  \left|{y_{23}^\ell  \over y_{33}^\ell } - {y_{23}^\nu \epsilon_4  \over y_{33}^\nu  \epsilon_5} \right| \geq   \left|{y_{23}^\ell  \over y_{33}^\ell }   \right| -   \left| {y_{23}^\nu   \over y_{33}^\nu  }  \right|   { \epsilon_4  \over  \epsilon_5}\approx  1 -  \sin \theta_{12}, \\ \nonumber
\sin \theta_{13}^\ell    &\simeq& \left|    -{y_{13}^\nu \epsilon_4  \over y_{33}^\nu  \epsilon_5} + {y_{13}^\ell   \epsilon_5  \over y_{33}^\ell  \epsilon_2 }    \right| \geq  \left|    - {y_{13}^\nu   \over y_{33}^\nu  }  \right|  { \epsilon_4  \over   \epsilon_5}-  \left| {y_{13}^\ell     \over y_{33}^\ell   }  \right| { \epsilon_5  \over   \epsilon_2}  \approx \sin \theta_{12} - \frac{m_s}{m_c},
\end{eqnarray}
where $ m_s /m_c =      \epsilon_5  /   \epsilon_2    $,.  Thus, we conclude that the leptonic mixing angles are predicted by  standard HVM in terms of the Cabibbo angle and masses of strange and charm quarks.  

We can now predict the leptonic mixing angles precisely using the measurement of the Cabibbo angle given in equation \ref{Qangle}.  For the first two mixing angles, we predict $\sin \theta_{12}^\ell =  0.55 \pm 0.00134 $  and $\sin \theta_{23}^\ell =  0.775 \pm 0.00067 $.  The  theoretical precision of the  $\sin \theta_{12}^\ell$  directly depends on the experimental precision of the Cabibbo angle, and   the theoretical precision of the  $\sin \theta_{23}^\ell$ is identical to that of the Cabibbo angle.  

The mixing angle $\sin \theta_{13}^\ell$ can be predicted using the values of running masses of strange and charm quarks  given at 1 TeV as $m_c = 0.532^{+0.074}_{-0.073}$ GeV and $m_s = 4.7^{+1.4}_{-1.3} \times 10^{-2}$ GeV\cite{Xing:2007fb}.  Using these values, we predict
$\sin \theta_{13}^\ell$ to be $ ( 0.1169 -0.1509)$.  However, we observe that the lower end of the experimental measurement  eliminates the range $(0.1169 - 0.1413)$  of the theoretical prediction.  On the other side, the upper end of the theoretical prediction rules out the range  $(0.1509 - 0.1550)$ of the experimental observation.  This leads to the final conclusive range of the angle  $\sin \theta_{13}^\ell = 0.1413 - 0.1509 $.

\subsection{Inverted mass ordering}
We note that neutrino masses may be inverted in order if the sign of the $\Delta m_{31}^2$ turns out to be negative. This scenario can be realised in the standard HVM by assigning the charges to  the right-handed neutrinos under the $\mathcal{Z}_2 \times \mathcal{Z}_4 \times \mathcal{Z}_{14} $ symmetry as : $ \nu_{e_R}: (-,  \omega,  \omega^{3} )$, $ \nu_{\mu_R}:  (-,  \omega,  \omega )$,  $\nu_{\tau_R}:  (+,  \omega,  \omega^{10} ) $ .  The mass matrix of the Dirac  neutrinos becomes,
\begin{align}
\M_{\rm Dirac} = \dfrac{v}{\sqrt{2}}  
\begin{pmatrix}
y_{11}^\nu   \epsilon_4 \epsilon_7   &  y_{12}^\nu  \epsilon_5  \epsilon_7  & 0 \\
y_{21}^\nu   \epsilon_4 \epsilon_7    & y_{22}^\nu  \epsilon_5  \epsilon_7 &  0 \\
y_{31}^\nu   \epsilon_2 \epsilon_7    &  y_{32}^\nu   \epsilon_4 \epsilon_7   &  y_{33}^\nu  \epsilon_1  \epsilon_7.
\end{pmatrix}.
\end{align} 
The masses of neutrinos are,
\begin{eqnarray}
\label{mass1b}
m_3  &\approx&  |y^\nu_{33}|  \epsilon_1 \epsilon_7 v/\sqrt{2},
m_2  \approx     \left|y^\nu_{22}  \right| \epsilon_5 \epsilon_7 v /\sqrt{2},
m_1  \approx   \left|y^\nu_{11} - \dfrac{y_{12}^\nu  y_{21}^\nu}{y_{33}^\nu} \right|\,  \epsilon_4 \epsilon_7 v /\sqrt{2}.
\end {eqnarray}
The  neutrino mixing angles now become,
\begin{eqnarray}
\sin \theta_{12}^\ell  &\simeq& \left|  {y_{12}^\ell   \epsilon_4  \over y_{22}^\ell  \epsilon_5 }  -{y_{12}^\nu  \over y_{22}^\nu }    \right| , 
\sin \theta_{23}^\ell  \simeq  \left|{y_{23}^\ell   \over y_{33}^\ell  }      \right| ,~
\sin \theta_{13}^\ell    \simeq \left|   {y_{13}^\ell   \epsilon_2  \over y_{33}^\ell  \epsilon_6 } \right| .
\end{eqnarray}  
In this case, our main predictions for the neutrino mixing angle are,
\begin{eqnarray}
\sin \theta_{12}^\ell  &\geq& 1 -  \sin \theta_{12}^q,~
\sin \theta_{13}^\ell    \geq    \frac{m_c}{m_b},
\end{eqnarray}
where  $\epsilon_7 \approx 10^{-10}$,   $  m_c / m_b =    \epsilon_2 /   \epsilon_6$.

We predict the range $\sin \theta_{13}^\ell \leq  0.18519 - 0.25958$ using  the mass of $b$-quark at 1 TeV  $m_b= 2.43\pm 0.08$ GeV\cite{Xing:2007fb}. The current global fit provides $\sin \theta_{13}^\ell = 0.14206 - 0.15569$.  We observe  that the equality sign of this  prediction is ruled out by the present data.  Thus, we conclude that  the inverted mass hierarchy of neutrinos is disfavoured in the SHVM.

\subsection{Majorana neutrinos}
There is a  possibility of the Majorana masses for neutrinos in this work.   We can  write the following Weinberg-type operators for the Majorana  mass,
\begin{eqnarray}
\label{mass6}
-{\mathcal{L}}_{\rm Weinberg}^{\ell} &=&   h_{ij}^\nu  \frac{\bar{\tilde{\psi}}_{L_i}^{\ell }   \varphi   \tilde{\varphi}^\dagger \psi_{L_j}^{\ell }}{\Lambda}   + h_{ij}^{\nu \prime }  \frac{\bar{\tilde{\psi}}_{L_i}^{\ell }   \varphi   \tilde{\varphi}^\dagger \psi_{L_j}^{\ell }}{\Lambda}   \left[  \dfrac{ \chi_i (\text{or}~ \chi_i^\dagger)}{\Lambda} \right] +
{\rm H.c.},
\end{eqnarray}
where $\tilde{\psi}_{L_i}^{\ell } = i \sigma_2 \psi_{L_i}^c  $.  It turns out that these operators are not allowed by the charge assignments given in table \ref{tab1}.  We may also have the right-handed neutrino mass operators $\mathcal{L}_{\rm M_R}$ given by,
\bea
\mathcal{L}_{\rm M_R}  &=& c_{ij}  \chi_i \bar{\nu_{f_R}^c} \nu_{f_R} + c_{ij}^\prime  \chi_i \bar{\nu_{f_R}^c} \nu_{f_R} \left[  \frac{\chi_i (\text{or}~ \chi_i^\dagger)}{\Lambda}  \right]+  c_{ij}^{\prime \prime}  \chi_i \bar{\nu_{f_R}^c} \nu_{f_R} \left[  \frac{\chi_i \chi_j (\text{or}~\chi_{i} \chi_{j}^\dagger)}{\Lambda^2}  \right]+
{\rm H.c.},
\eea
However, they are forbidden in the standard HVM. Therefore, neutrinos in the standard HVM,  are   Dirac type.

\section{Scalar potential of the SHVM}
\label{scalar_potential}
The scalar potential is written as,
\bea 
V &=& \mu^2 \varphi^\dagger \varphi + \lambda (\varphi^\dagger \varphi)^2  +   \mu_{\chi_1}^2 |\chi_1|^2   +   \mu_{\chi_2}^2 |\chi_2|^2   +   \mu_{\chi_3}^2 |\chi_3|^2    +   \mu_{\chi_4}^2|\chi_4|^2    +   \mu_{\chi_5}^2 |\chi_5|^2    +   \mu_{\chi_6}^2 |\chi_6|^2    
 \\ \nonumber  
 &&+   \mu_{\chi_7}^2 |\chi_7|^2 
+ \lambda_{\chi_1}  |\chi_1|^4  + \lambda_{\chi_2}  |\chi_2|^4+ \lambda_{\chi_3}  |\chi_3|^4+ \lambda_{\chi_4}  |\chi_4|^4+ \lambda_{\chi_5}  |\chi_5|^4+ \lambda_{\chi_6}  |\chi_6|^4+ \lambda_{\chi_7}  |\chi_7|^4   \\ \nonumber 
&&+     \lambda_{\varphi \chi}  \varphi^\dagger \varphi   \chi^\dagger_i \chi_j  + \lambda_{\chi_{12}}  |\chi_1|^2 |\chi_2|^2  + \lambda_{\chi_{13}}  |\chi_1|^2 |\chi_3|^2  + \lambda_{\chi_{14}}  |\chi_1|^2 |\chi_4|^2  + \lambda_{\chi_{15}}  |\chi_1|^2 |\chi_5|^2   \\ \nonumber
&&+ \lambda_{\chi_{16}}  |\chi_1|^2 |\chi_6|^2  + \lambda_{\chi_{17}}  |\chi_1|^2 |\chi_7|^2 
+ \lambda_{\chi_{23}}  |\chi_2|^2 |\chi_3|^2  + \lambda_{\chi_{24}}  |\chi_2|^2 |\chi_4|^2  + \lambda_{\chi_{25}}  |\chi_2|^2 |\chi_5|^2  + \lambda_{\chi_{26}}  |\chi_2|^2 |\chi_6|^2   \\ \nonumber
&&+ \lambda_{\chi_{27}}  |\chi_2|^2 |\chi_7|^2  
+ \lambda_{\chi_{34}}  |\chi_2|^3 |\chi_3|^4  + \lambda_{\chi_{35}}  |\chi_2|^3 |\chi_5|^2  + \lambda_{\chi_{36}}  |\chi_3|^2 |\chi_6|^2  + \lambda_{\chi_{37}}  |\chi_3|^2 |\chi_7|^2   \\ \nonumber
&&+ \lambda_{\chi_{45}}  |\chi_4|^2 |\chi_5|^2  + \lambda_{\chi_{46}}  |\chi_4|^2 |\chi_6|^2  + \lambda_{\chi_{47}}  |\chi_4|^2 |\chi_7|^2 + \lambda_{\chi_{56}}  |\chi_5|^2 |\chi_6|^2  + \lambda_{\chi_{57}}  |\chi_5|^2 |\chi_7|^2  \\ \nonumber
&&  + \lambda_{\chi_{67}}  |\chi_6|^2 |\chi_7|^2  + \lambda_{1122}  \chi_1^2 \chi_2^2 + \lambda_{1133}  \chi_1^2 \chi_3^{\dagger 2}  + \lambda_{2233}  \chi_2^2 \chi_3^2 
+ \lambda_{2224}  \chi_2^3 \chi_4  + \lambda_{4436}  \chi_4^2 \chi_3^{\dagger } \chi_6^{\dagger } + \lambda_{2555}  \chi_2 \chi_5^{\dagger 3}  \\ \nonumber
&& + \sigma_{16} \chi_1^2 \chi_6  + \sigma_{26} \chi_2^2 \chi_6^\dagger + \sigma_{36} \chi_3^2 \chi_6  + \sigma_{43} \chi_4^2 \chi_3 + \sigma_{73} \chi_7^2 \chi_3^\dagger 
+ \sigma_{345} \chi_3 \chi_4^\dagger \chi_5  + \sigma_{235} \chi_2 \chi_3^\dagger \chi_5   \\ \nonumber
&& + \sigma_{456} \chi_4  \chi_5 \chi_6^\dagger  + \sigma_{246} \chi_2  \chi_4 \chi_6 + \sigma_{345} \chi_3  \chi_4^\dagger \chi_5
+ \rm H.c..
\label{SP} 
\eea

We can parametrize the scalar fields  as,
\begin{align}
 \chi_i(x) &=\frac{v_i + s_i(x) +i\, a_i(x)}{\sqrt{2}}, ~ \varphi =\left( \begin{array}{c}
G^+ \\
\frac{v+h+i G^0}{\sqrt{2}} \\
\end{array} \right).
\end{align}

\subsection{Emergence of  a multiple ALPs scenario}
The pseudoscalars $a_i$ can achieve their masses through the potential \cite{Abbas:2023ion},
\begin{equation} \label{VN}
 V_\chi  = -\lambda_{i}^\prime {\chi_i^{\tilde N} \over \Lambda^{\tilde N-4}} + \text{H.c.}, 
\end{equation}
where $\tilde N $ is the least common multiple of the $\rm N$, $\rm M$ and $\rm P$ in the  $\mathcal{Z}_{\rm N} \times \mathcal{Z}_{\rm M} \times \mathcal{Z}_{\rm P}$ flavour symmetry. The masses of pseudo-scalar particles are now given by,

\begin{equation} 
\label{mai}
 m_{a_i}^2={1\over8} |\lambda_{i}^\prime| \tilde N^2 \epsilon_i^{\tilde N-4} v_i^2.
\end{equation}

From these,  we can calculate the masses of the pseudoscalars $a_i$.  For instance, for $|\lambda^\prime| = 1$  and $\tilde N = 14$, the masses are given for three different values of $v_i$ in table \ref{tab:tabai}.   The $\tilde N = 14$ corresponds to the symmetry $\mathcal{Z}_2 \times \mathcal{Z}_{\rm M} \times \mathcal{Z}_{14} $  where $\rm M= 7 ~\text{or}~ 14$.  Thus, we see that except the pseudoscalar $a_3$, all the other pseudoscalars are are very light belwo the scale $10^{11}$ GeV, and are acting like ALPS.  The pseudoscalar $a_6$ becomes heavier at the scale above $10^{11}$ GeV.

We notice that the pseudoscalars $a_{4,5,6}$ contribute to the neutral meson mixing ($K,B_d, B_s$) parameters above the scale of the mixing processes.  However,  the contribution of the pseudoscalars $a_{4,5,6}$ to the neutral meson mixing ($K,B_d, B_s$)  parameters is proportional to $1/v_i^2 $.  Therefore,  this contribution is hugely suppressed,  and does not enhance the SM contribution \cite{Abbas:2024jut}.  Thus,  there are no observable effects of the particles $a_{4,5,6}$ in the quark and leptonic flavor physics.  However,  the pseudoscalar $a_3$ may reveal itself through the top-quark interactions such as the decays $ a_3 \rightarrow \gamma \gamma, t\bar{t} $  providing a smoking gun signature of the SHVM \cite{Abbas:2024jut}.

\begin{table}[ht]
\begin{center}
\resizebox{\textwidth}{!}{
\begin{tabular}{|c|c|c|c|c|c|c|c|}
  \hline
  Pseudoscalars                               &   $a_1$ &  $a_2$   &  $a_3$  & $a_4$   & $a_5$   &  $a_6$ &  $a_7$    \\
  \hline
  \hline
   Mass (GeV)         &      &         &   &   &       &      &         \\
    at $v_i = 10^{2}$ GeV        &    $1.6  \times 10^{-25}$    &    $1.4  \times 10^{-10}$        &  $ 246.71$   & $ 4.18  \times 10^{-19}$   &  $ 7.10  \times 10^{-16}$     &  $ 2.27  \times 10^{-9}$       &  $ 9.45  \times 10^{-44}$        \\
   Mass (GeV)         &      &         &   &   &       &      &         \\
    at $v_i = 10^{11}$ GeV        &    $1.6 \times 10^{-16}$    &   $ 0.14$          &  $ 2.47  \times 10^{11}$   & $ 4.18  \times 10^{-10}$      &  $ 7.10  \times 10^{-7}$      &  $ 2.27$      &   $ 9.45  \times 10^{-35}$         \\
  Mass (GeV)         &      &         &   &   &       &      &         \\
    at $v_i = 10^{19}$ GeV        &    $1.6 \times 10^{-8}$    &   $1.42 \times 10^7$          &  $ 2.47  \times 10^{19}$   & $0.04$    &  $ 71.02$      &  $ 2.27 \times 10^8$      &   $ 9.45  \times 10^{-27}$  \\
   \hline
     \end{tabular}}
\end{center}
\caption{The masses of pseudoscalars $a_i$ for $|\lambda^\prime| = 1$ and $\tilde N = 14$.}
 \label{tab:tabai}
\end{table}

\subsection{Masses of scalars}
We assume    $  \lambda_{\varphi \chi}  =0$ and $ \lambda_{\chi_i} =\lambda_\chi$  for simplicity, and diagonalize the the mass matrix square of the scalar particles numerically  by writing $v_i = \sqrt{2} \epsilon_i \Lambda$ where $\epsilon_i$ can be found  in equation \ref{epsi}.  Thus, we have,

\begin{align}
    U^T \mathcal{M}_s^2 U= \mathcal{M}_{dia}^2,
\end{align}

where $\mathcal{M}_s^2$ denotes the mass matrix square of the scalars, $\mathcal{M}_{dia}$ shows the diagonalized scalar mass matrix, and $U$ represents an orthogonal matrix.  These matrices can be found  in appendix.

The physical scalar states can be defined by the following transformation,

    \begin{eqnarray}
     \begin{pmatrix}
s_7 \\
s_1 \\
s_4  \\
s_5 \\
s_2 \\
 s_6  \\
s_3\\
\end{pmatrix} = U^T
  \begin{pmatrix}
h_7 \\
h_1 \\
h_4  \\
h_5 \\
h_2 \\
 h_6  \\
h_3\\
\end{pmatrix} , 
\label{scalar_trans}    
\end{eqnarray}

The masses of scalars turn out to be,
\begin{align}
    m_{h_7}^2  & = 4.72 \times 10^{-18} \lambda_\chi \Lambda^2, ~ m_{h_1}^2  = 9.33 \times 10^{-11} \lambda_\chi \Lambda^2,~ m_{h_4}^2  = 3.57 \times 10^{-8} \lambda_\chi \Lambda^2 \\ \nonumber
      m_{h_5}^2 & = 7.3 \times 10^{-7} \lambda_\chi \Lambda^2,~ m_{h_2}^2  = 9.45 \times 10^{-5} \lambda_\chi \Lambda^2,~  m_{h_6}^2  = 3.67 \times 10^{-4} \lambda_\chi \Lambda^2,~ m_{h_3}^2  = 12 \lambda_\chi  \Lambda^2
\end{align}

\section{A possible dark matter candidate}
\label{NDM}
The SHVM framework may provide a new class of dark matter candidate named as ``neutrinic dark matter".  This can be found by observing that if the ALPS  $a_{i}$ are dark matter candidate, the most powerful  bounds on the scale  $\Lambda$ come from the FCNC process $K^+ \to \pi^+ a_{4,5}$.  The FCNC process $K^+ \to \pi^+ a_{4,5}$   implies that  $ v_{4,5} \gtrsim 7 \times 10^{11} V^d_{21} {\rm GeV}$ where  $V^d_{21} \approx 0.225$ \cite{Bjorkeroth:2018dzu}. This corresponds to $\Lambda = v_i/\sqrt{2} \epsilon_i \approx 10^{17}$ GeV.   We do not consider this scenario since it will lead to the extreme suppression of the all quark and leptonic FCNC processes of the SHVM.   Thus, there will be no signature of the SHVM in flavour and collider physics.  Hence, we do not consider the pseudoscalar $a_{i} ($i=1-6$)$ to be ALPs and dark matter candidates.  This can be achieved by making  the pseudoscalars $a_{i} ($i=1-6$)$  very heavy through the following soft symmetry breaking potential.

\begin{align}
V_{\rm soft}^{a_i} =  \rho_i (\chi_i^2 + \chi_i^{*2}),
\end{align}
where $i=1-6$.  This term can be generated by the strong dynamics of the dark-technicolour, whose the fields $\chi_i$ ($i=1-6$) are the bound states,  discussed in the appendix.  The mass of the pseudoscalars $a_{i}$ now reads,

\begin{align}
m_{a_i} = \sqrt{\rho_i},
\end{align}
where $i=1-6$.  Now the masses $m_{a_i}$ are free parameters, they can be anywhere between the eletroweak and the Planck scale.

The only particle which maybe a dark matter candidate is the pseudoscalar $a_7$.  This is a bound state of the dark-QCD in the dark technicolour model as discussed in the appendix.  The interaction between the dark  technicolour whose bound states are $\chi_i$ ($i=1-6$) and the dark-QCD are extremely suppressed in the UV completion.  Therefore, we assume that  there is zero interaction, effectively, between the scalar fields $\chi_i$ and $\chi_7$. For the VEV $v_7 = 1$ GeV, the scale $\Lambda$ should be of the order $10^9$ GeV.  The mass of the ALP $a_7$ from equation \ref{mai} is $\approx 10^{-38}$ eV for  $\lambda_{7}^\prime=1$ and $\lambda = 10^9$ GeV.  The lower bound on the mass of cold dark matter particle is $10^{-21}$ eV \cite{Cirelli:2024ssz}.  We observe that for the scale $\Lambda = 10^{19}$ GeV, the mass of the ALP  $a_7$ is of the order $10^{-26}$ eV.  Thus, the mass of the ALP  $a_7$ is far below the lower bound of the dark matter.  Therefore, the ALP   $a_7$ cannot be a dark matter candidate for the mass given by equation  \ref{mai}.

However, the ALP   $a_7$ may be a  novel class of   dark matter  if we introduce its mass through the soft symmetry breaking potential given by,
\begin{align}
V_{\rm soft}^{a_7} =  \rho_7 (\chi_7^2 + \chi_7^{*2}).
\end{align}

The mass of the ALP $a_7$ is now given as,
\begin{align}
m_{a_7} = \sqrt{\rho_7},
\end{align}
which is a free parameter.  The only decay channel of the ALP $a_7$ is to a pair of neutrinos.   The decay rate is,

\begin{eqnarray}
 \Gamma (a_7 \to \nu \nu) =  \frac{1}{8\pi} g_{a_7 \nu\nu}^2 m_{a_7} \beta_\nu,
  \end{eqnarray}
  
where $\beta_f = (1-4 m_\nu^2/m_{a_7}^2)^{1/2}$ and $g_{a_7\nu\nu} =  \frac{v}{ v_7 \sqrt{2}} \epsilon_{5} \epsilon_7 =    \frac{v}{ 2 \Lambda} \epsilon_{5}$. For  $m_{a_7} = 10^{-8}\rm GeV$, the $\Gamma (a_7 \to \nu \nu) = 8.8 \times 10^{-45} $ GeV where  the scale $\Lambda $ is $ 10^{16}$ GeV.  However,  the scale $\Lambda$ can be as low as the eletroweak scale if  $m_{a_7}< 2 m_\nu$.  In this scenario,  $a_7$ may still be a possible dark-matter particle if its mass is greater than  $10^{-21}$ eV \cite{Cirelli:2024ssz}.
  
The possible neutrinic dark matter may reveal itself either through the Higgs-$\chi_7$ portal (the term $  \varphi^\dagger \varphi   \chi^\dagger_7 \chi_7 $),  or $\chi_i$-$\chi_7$  portal (the term $   \chi_i^\dagger \chi_i   \chi^\dagger_7 \chi_7 $).  The crucial difference between the possible neutrinic dark matter and other ALP dark matter is that the possible neutrinic dark matter does not have coupling to electromagnetic field at leading-order.  Therefore, the operator $\frac{1}{4} g_{a_7 \gamma\gamma} \varphi  F^{\mu\nu} \Tilde{F}_{\mu \nu}$ effectively does not exist.  This keeps the possible neutrinic dark matter apart from other ALPs dark matter candidates.

\section{Summary}
\label{sum}
In this work, we have derived a framework to address, in particular,  the leptonic  flavour mixing based on the VEVs hierarchy.  We name it the standard HVM.  The standard HVM predicts its unique and defining signatures in the form of inequalities $\sin \theta_{12}^\ell   \geq 1 - 2 \sin \theta_{12}$,  $\sin \theta_{23}^\ell   \geq 1 -  \sin \theta_{12}$, and $\sin \theta_{13}^\ell  \geq \sin \theta_{12} - \frac{m_s}{m_c}$ for the normal mass ordering.  This results in very precise predictions of the three leptonic mixing angles given by  $\sin \theta_{12}^\ell =  0.55 \pm 0.00134 $,  $\sin \theta_{23}^\ell =  0.775 \pm 0.00067 $,  and   $\sin \theta_{13}^\ell = 0.1413 - 0.1509 $.   Inverted neutrino mass ordering is disfavoured by the SHVM.

The next generation neutrino experiments such as DUNE, Hyper-Kamiokande, and JUNO \cite{Huber:2022lpm} may  probe these predictions.  Furthermore, we have predicted the   quark mixing angle $\sin \theta_{13}$  to be $  \approx \sin \theta_{12}^{ 3} - 2 \frac{m_c}{m_t} $, which is perfectly  in agreement with the observed value.  However, for this prediction, we need to accept a reasonable assumption that $| y_{13}^u  / y_{33}^u   | \approx 2 $.  An important implication of the SHVM is a new class of possible dark matter candidate, named as neutrinic dark matter. The flavour physics and collider signatures  of the SHVM are investigated in reference \cite{Abbas:2024jut}.

\section*{Acknowledgement}
Author is grateful to  the reviewer for pointing out  important missing terms in the scalar potential.  The Feynman diagrams  are created by the {\tt JaxoDraw} \cite{Binosi:2008ig}. This work is supported by the  Council of Science and Technology,  Govt. of Uttar Pradesh,  India through the  project ``   A new paradigm for flavour problem "  no.   CST/D-1301,  and Anusandhan National Research Foundation (SERB),  Department of Science and Technology, Government of India through the project `` Higgs Physics within and beyond the Standard Model" no. CRG/2022/003237. 

\section*{Appendix}

\subsection*{Benchmark couplings}
For reproducing the fermion masses and mixing from the standard HVM,  the values of the fermion masses at $ 1$TeV,  given in ref. \cite{Xing:2007fb}, are used.  The quark mixing parameters are taken from ref.  \cite{pdg22}.  The neutrino  oscillation data for the normal hierarchy are taken   from the global fit results given in ref.  \cite{deSalas:2020pgw}.   The coefficients $y_{ij}^{u,d,\ell,\nu}= |y_{ij}^{u,d,\ell,\nu}| e^{i \phi_{ij}^{q,\ell,\nu}}$  are scanned  in the ranges $|y_{ij}^{u,d,\ell, \nu}| \in [0.9,3]$ and $ \phi_{ij}^{q,\ell,\nu} \in [0,2\pi]$.  The fitting results are,
\begin{equation*}
y^u_{ij} = \begin{pmatrix}
1.0 + 1.73 i& 0 & 0.58 + 1.95 i  \\
0 & -0.89 + 0.45 i &  -0.98 - 0.14 i \\
0 & -0.32 + 1.60 i &  1 - 0.02 i
\end{pmatrix},  
\end{equation*}

\begin{equation*}
y^d_{ij} = \begin{pmatrix}
-1.77+2.19 i & -1.30 +2.29 i & 1.08\, -2.16 i   \\
0.76\, -0.95 i & 1.12\, -2.36 i & 2.09\, +0.24 i \\
0.31\, +1.98 i & -2.25+0.16 i &  -1.65-2.04 i
\end{pmatrix},  
\end{equation*}
The Dirac CP phase in  the standard parametrization is $\delta_{\rm CP}^q \approx 1.144$.

\begin{equation*}
y^\ell_{ij} = \begin{pmatrix}
1.64\, -0.20 i & 0.52\, -0.89 i & 0.36\, -1.04 i \\
0  & 0.07\, +1.01 i & 0.42\, +1.02 i \\
0 & 0 & 1.09
\end{pmatrix},
\end{equation*}

For normal mass ordering the neutrino couplings are,
\begin{equation*}
y^\nu_{ij} = \begin{pmatrix}
0.6\, -0.88 i & 0.87\, -0.41 i & 1.5\, -0.00004 i \\
0 & 0.96\, +0.12 i & -0.52-1.41 i \\
0 & -1.43+0.28 i & 1.5\, +0.00004 i
\end{pmatrix}, 
\end{equation*}
and the leptonic Dirac $CP$ phase  is $\delta_{\rm CP}^\ell  \approx  3.14$.

\subsection*{Scalar mass matrix diagonalization}
 For simplification, we impose an extra $\mathcal{Z}_2^\prime$ symmetry on the right-handed fermions such that right-handed  fermions transform as  $\psi_R : -$, the singlet scalar fields $\chi$ as  $\chi_i: -$ where $i=1-6$, and the field $\chi_7$ as $\chi_7:+$ under this new  $\mathcal{Z}_2^\prime$ symmetry.  This will forbid all cubic terms in the scalar potential without affecting the flavour structure of the model.

\begin{equation*}
m_{s}^2 = \lambda_{\chi} \Lambda^2 \begin{pmatrix}
16 \epsilon_{7}^2  & 8 \epsilon_{1} \epsilon_{7}  & 8 \epsilon_{4} \epsilon_{7}
    & 8 \epsilon_{5} \epsilon_{7}  & 8 \epsilon_{2} \epsilon_{7}  & 8 \epsilon_{6}\epsilon_{7} & 8 \epsilon_{3} \epsilon_{7}  \\
 8 \epsilon_1 \epsilon_{7}  & 16 \epsilon_1^2  & 8 \epsilon_1 \epsilon_{4}
   & 8 \epsilon_1 \epsilon_{5}  & 8 \epsilon_1 \epsilon_{2}  & 8 \epsilon_1 \epsilon_{6} & 8 \epsilon_1 \epsilon_{3}  \\
 8 \epsilon_{4} \epsilon_{7}  & 8 \epsilon_1 \epsilon_{4}  & 16 \epsilon_{4}^2
   & 8 \epsilon_{4} \epsilon_{5}  & 8 \epsilon_{2} \epsilon_{4} & 8 \epsilon_{4} \epsilon_{6} & 8 \epsilon_{3} \epsilon_{4}  \\
 8 \epsilon_{5} \epsilon_{7}  & 8 \epsilon_1 \epsilon_{5}  & 8 \epsilon_{4}
   \epsilon_{5}  & 16 \epsilon_{5}^2  & 8 \epsilon_{2} \epsilon_{5}  & 8 \epsilon_{5} \epsilon_{6} & 8 \epsilon_{3} \epsilon_{5} \\
 8 \epsilon_{2} \epsilon_{7}  & 8 \epsilon_1 \epsilon_{2}  & 8 \epsilon_{2}
   \epsilon_{4}  & 8 \epsilon_{2} \epsilon_{5}  & 16 \epsilon_{2}^2  & 8 \epsilon_{2} \epsilon_{6} & 8 \epsilon_{2} \epsilon_{3}  \\
 8 \epsilon_{6}\epsilon_{7} & 8 \epsilon_1 \epsilon_{6} & 8 \epsilon_{4}
   \epsilon_{6} & 8 \epsilon_{5} \epsilon_{6} & 8 \epsilon_{2} \epsilon_{6}
   & 16 \epsilon_{6}^2  & 8 \epsilon_{3} \epsilon_{6} \\
 8 \epsilon_{3} \epsilon_{7}  & 8 \epsilon_1 \epsilon_{3}  & 8 \epsilon_{3}
   \epsilon_{4}  & 8 \epsilon_{3} \epsilon_{5}  & 8 \epsilon_{2} \epsilon_{3} & 8 \epsilon_{3} \epsilon_{6} & 16 \epsilon_{3}^2  \\
\end{pmatrix},
\end{equation*}

\begin{equation*}
U =\begin{pmatrix}
1 & -3.24 \times 10^{-4} & -1.94 \times 10^{-6} & -5.34 \times 10^{-7} & -4.88 \times 10^{-8} & 4.69 \times 10^{-8} & 4.14 \times 10^{-10} \\
    -3.24 \times 10^{-4} & -0.999 & -8.56 \times 10^{-3} & -2.35 \times 10^{-3} & -2.15 \times 10^{-4} & 2.06 \times 10^{-4} & 1.82 \times 10^{-6} \\
    -1.68 \times 10^{-6} & 8.66 \times 10^{-3} & -0.999 & -4.73 \times 10^{-2} & -4.18 \times 10^{-3} & 3.98 \times 10^{-3} & 3.52 \times 10^{-4} \\
    -3.79 \times 10^{-7} & 1.95 \times 10^{-2} & 4.75 \times 10^{-2} & -0.998 & -1.85 \times 10^{-2} & 1.77 \times 10^{-2} & 1.56 \times 10^{-4} \\
    -3.37 \times 10^{-8} & 1.73 \times 10^{-4} & 3.95 \times 10^{-3} & 2.25 \times 10^{-2} & -0.968 & 2.49 \times 10^{-1} & 1.76 \times 10^{-3} \\
    -1.89 \times 10^{-8} & 9.75 \times 10^{-5} & 2.23 \times 10^{-3} & 1.26\times 10^{-2} & 2.499 \times 10^{-1} & 0.97 & 3.12 \times 10^{-3} \\
    -1.18 \times 10^{-10} & 6.08 \times 10^{-7} & 1.39 \times 10^{-5} & 7.84 \times 10^{-4} & 9.27 \times 10^{-4} & -3.46 \times 10^{-3} & 0.999
\end{pmatrix}
\end{equation*}

\begin{equation*}
M_{dia}^2 =  \lambda_{\chi} \Lambda^2 
  \text{dia} (4.71 \times 10^{-18}, 9.33 \times 10^{-11}, 3.56\times10^{-8},  7.30 \times 10^{-7}, 9.45 \times 10^{-5}, 3.67 \times 10^{-4}, 12.01). \\
\end{equation*}

\subsection{UV completion in the dark-QCD paradigm: A minimal model}
A UV completion of the SHVM is possible in the DTC paradigm discussed in reference \cite{Abbas:2020frs}. In this section, we propose a new dark-QCD (DQCD) paradigm, which provides a UV completion in a minimal framework.  The DQCD paradigm is based on a minimal symmetry  $\mathcal{G}_{DQCD} = SU(\rm{N}_{\rm{TC}}) \times  SU(\rm{N}_{\rm{DQ}})  $, where   $\rm DQ$ shows   the DQCD dynamics of vector-like fermions.   The difference between the DTC and the DQCD paradigm is the absence of the $SU(\rm N_{\rm DTC})$ dynamics in the later.  The multi-fermions chiral condensates, which play the role of the VEVs $\langle \chi_r \rangle$,  are formed by the TC fermions $ D_{L,R}$.

The transformations of TC fermions  under the $SU(3)_c \times SU(2)_L \times U(1)_Y \times \mathcal{G}_{DQCD}$ is,
\begin{eqnarray}
T^i  &\equiv&   \begin{pmatrix}
T  \\
B
\end{pmatrix}_L:(1,2,0,\rm{N}_{\rm TC},1), ~
T_{R}^i : (1,1,1,\text{N}_{\rm{TC}},1), B_{R}^i : (1,1,-1,\rm{N}_{\rm TC},1), \\ \nonumber
 D_{L,R}  &\equiv& C_{i,  L,R}
 : (1,1,1,\text{N}_{\rm{TC}},1),  
  S_{i, L,R}  
 : ( 1,1,-1,\text{N}_{\rm{TC}},1),   
\end{eqnarray}
where $i=1,2,3 \cdots $, and  electric charges $+\frac{1}{2}$ for $T$ and $\mathcal C$ ,  and $-\frac{1}{2}$ for $B$ and $\mathcal S$. 

The  $ SU(\rm N_{\rm DQ})$ dynamics has vector-like fermions transforming as,
\begin{eqnarray}
F_{L,R} &\equiv &U_{L,R}^i \equiv  (3,1,\dfrac{4}{3},1, \rm{N}_{\rm F}),
D_{L,R}^{i} \equiv   (3,1,-\dfrac{2}{3},1,\rm{N}_{\rm F}),  \\ \nonumber 
N_{L,R}^i &\equiv&   (1,1,0,1,\rm{N}_{\rm F}), 
E_{L,R}^{i} \equiv   (1,1,-2,1,\rm{N}_{\rm F}).
\end{eqnarray}

The multi-fermion condensate is paramerized as  \cite{Aoki:1983yy}, 
\be 
\label{VEV_h}
\langle  ( \bar{\psi}_L \psi_R )^n \rangle \sim \left(  \Lambda \exp(k \Delta \chi) \right)^{3n},
\ee
where $\Delta \chi$ is the chirality of the corresponding operator, $k$ is a constant, and $\Lambda$ denotes  the scale of the  gauge dynamics.  

In this model, we have three axial $U(1)_A^{\rm QCD, TC,   DQCD}$ symmetries.  They are broken to  a  discrete cyclic group by instantons as  \cite{Harari:1981bs},
\begin{equation}
 U(1)_A^{\rm QCD, TC,  DQCD} \rightarrow \mathcal{Z}_{2 \rm K_{\rm QCD, TC,  DQCD}},
\end{equation}
where $\rm K_{\rm QCD, TC,  DQCD} $ are number of  fundamental massless flavours of the QCD,  TC,  and DQCD gauge dynamics in the $N$-dimensional representation of the gauge group $SU(\rm N)_{\rm QCD, TC,  DQCD}$.   This results in  a generic   $\mathcal{Z}_{\rm N} \times \mathcal{Z}_{\rm M} \times \mathcal{Z}_{\rm P}$ flavour symmetry, where $\rm N= 2 \rm K_{\rm QCD}$, $\rm M= 2 \rm K_{\rm TC}$ and $\rm P= 2 \rm K_{\rm DQCD}$. 

\begin{figure}[h]
	\centering
 \includegraphics[width=\linewidth]{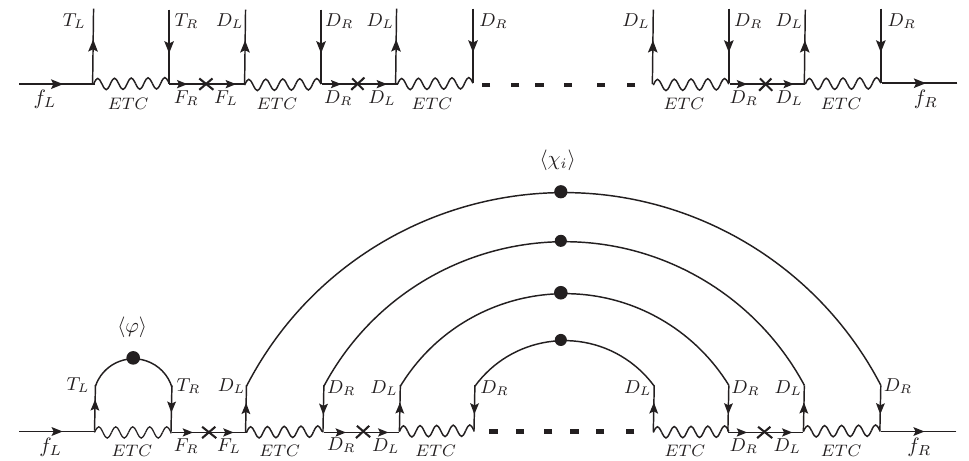}
    \caption{The Feynman diagrams for the masses of the  charged fermions in the  DQCD paradigm.}
 \label{fig1}	
 \end{figure}

We assume that TC  and the SM fermions can be accommodated in an extended TC sector (ETC).  In the DQCD paradigm,  the masses of fermions of SHVM originate from the generic interactions given in figure \ref{fig1} on the top.  The formation of multi-fermion chiral consdensates (denoted by blob) and resulting  mass of the SM fermion is shown below  in figure \ref{fig1}.

We write the charged fermion masses as,
\bea
\label{TC_masses}
m_{f} & = & y_f  \frac{\Lambda_{\text{TC}}^{3}}{\Lambda_{\text{ETC}}^2}  \dfrac{1}{\Lambda_{\text{F}}} \frac{\Lambda_{\text{TC}}^{n_i + 1}}{\Lambda_{\text{ETC}}^{n_i}} \left[ \exp(n_i k) \right]^{n_i/2},~
\eea
where $n_i = 2,4, \cdots 2 n $ denotes the number of fermions in a multi-quark chiral condensate which play the role of the VEV $ \langle \chi_i \rangle$ in the SHVM. 

The  form of  the $\epsilon_i$ parameters appearing in the SHVM can be written as,
\bea
\epsilon_i \propto  \dfrac{1}{\Lambda_{\text{F}}} \frac{\Lambda_{\text{DTC}}^{n_i + 1}}{\Lambda_{\text{EDTC}}^{n_i}}  \left[ \exp(n_i k) \right]^{n_i/2}.
\eea

\begin{figure}[h]
	\centering
 \includegraphics[width=\linewidth]{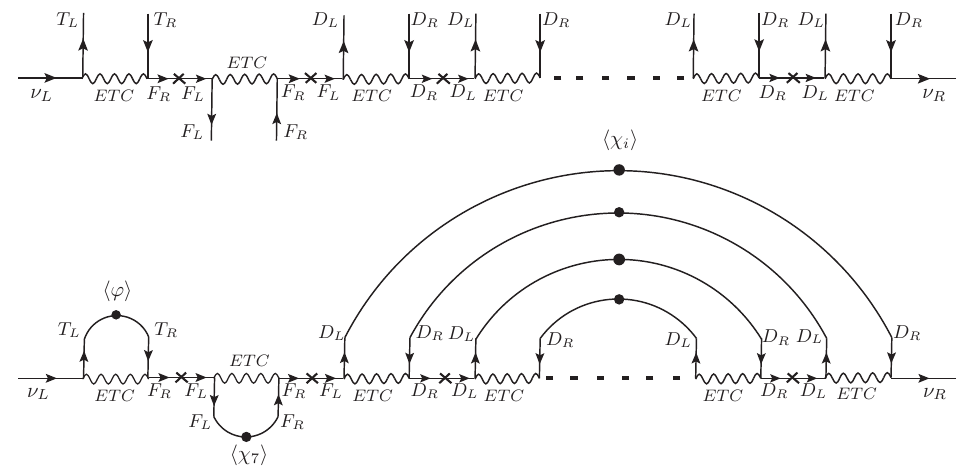}
    \caption{The Feynman diagrams generating dimension-6 operators, responsible for neutrino masses,  given in equation \ref{neutrino_mass}.    On the top, the generic interactions involving the SM, TC, DTC gauge sectors mediated by ETC are shown.  In the bottom, we show the formation of the fermionic condensate $\langle \bar{F}_L F_R \rangle$ which is the VEV $\langle \chi_7 \rangle$ in the SHVM. }
 \label{fig2}	
 \end{figure}

The creation of the dimension-6 operators responsible for neutrino masses,  given in equation \ref{neutrino_mass}, is shown in figure \ref{fig2}.   The chiral condensate  $\langle \bar{F}_L F_R \rangle$ plays the role of the VEV $\langle \chi_7 \rangle$.

The neutrinos masses are recovered as,
\bea
\label{neutrino_mass}
m_{\nu} & = & y_f  \frac{\Lambda_{\text{TC}}^{3}}{\Lambda_{\text{ETC}}^2}  \dfrac{1}{\Lambda_{\text{F}}} \frac{\Lambda_{\text{DTC}}^{n_i + 1}}{\Lambda_{\text{ETC}}^{n_i}} \exp(n_i k)  \dfrac{1}{\Lambda_{\text{F}}} \frac{\Lambda_{\text{F}}^{3}}{\Lambda_{\text{ETC}}^{2}} \exp(2 k),
\eea
where,  
\bea
\epsilon_7 \propto  \dfrac{1}{\Lambda_{\text{F}}} \frac{\Lambda_{\text{F}}^{3}}{\Lambda_{\text{ETC}}^{2}} \exp(2 k).
\eea

\end{document}